\documentstyle[pra,twocolumn,aps,epsf,floats]{revtex}
\begin{document}
\draft\narrowtext
\title{Tomographic measurement of nonclassical radiation states}
\author{G. M. D'Ariano and M. F. Sacchi}
\address{
Theoretical Quantum Optics Group\\
Universit\`a degli Studi di Pavia, INFM --- Unit\`a di Pavia, 
via A. Bassi 6, I-27100 Pavia, Italy}
\author{Prem Kumar}
\address{Department of Electrical and Computer Engineering\\
Northwestern University, 2145 North Sheridan Road, Evanston, 
IL 60208-3118, USA}
\maketitle
\begin{abstract} 
We propose to experimentally test the nonclassicality of quantum
states through homodyne tomography. For single-mode states we check
violations of inequalities involving the photon-number
probability. For two-mode states we test the nonclassicality by
reconstructing some suitable number-operator functions.  The test can
be performed with available quantum efficiency of homodyne detection,
by measuring the pertaining quantities on the corresponding noisy
states.
\end{abstract}
\pacs{1998 PACS number(s): 03.65.-w, 03.65.Bz, 42.50.Dv, 42.50-p}
\nopagebreak
\section{INTRODUCTION} 
The concept of nonclassical states of light has drawn much
attention in quantum optics
\cite{mand,mand2,hill,sch,spe,tomb,agar,lee1,kly1,iss,jpa}.  The
customary definition of nonclassicality is given in terms of the
Glauber-Sudarshan $P$-function: a nonclassical state does not admit a
regular positive $P$-function representation, namely, it cannot be
written as a statistical mixture of coherent states. Such states
produce effects that have no classical analogue.  These kinds of states
are of fundamental relevance not only for the demonstration of the
inadequacy of classical description, but also for applications, e.g.,
in the realms of information transmission and interferometric
measurements \cite{spe,tomb,iss}.  \par In this paper we are
interested in testing the nonclassicality of a quantum state by means
of an operational criterium, which is based on a set of quantities
that can be measured experimentally with some given level of
confidence, even in the presence of loss, noise, and less-than-unity
quantum efficiency. The positivity of the $P$-function itself cannot
be adopted as a test, since there is no method available to measure
it. The $P$-function is a Fourier transform on the complex plane of
the generating function for the normal-ordered moments; hence, in
principle, it could be recovered by measuring all the quadrature
components of the field, and subsequently performing an (deconvolved)
inverse Radon transform \cite{vog}.  Currently, there is a
well-established quantitative method for such a universal homodyne
measurement, and it is usually referred to as quantum homodyne
tomography (see Ref.~\cite{bilk} for a review). However, as proven in
Ref.~\cite{univ}, only the generalized Wigner functions of order
$s<1-\eta ^{-1}$ can be measured, $\eta $ being the quantum efficiency
of homodyne detection. Hence, through this technique, all functions
from $s=1$ to $s=0$ cannot be recovered, i.e., we cannot obtain the
$P$-function and all its smoothed convolutions up to the customary
Wigner function.  For 
the same reason, the nonclassicality parameter
proposed by Lee \cite{lee1}, namely, the maximum $s$-parameter that
provides a positive distribution, cannot be experimentally measured.
\par Among the many manifestations of nonclassical effects, one finds
squeezing, antibunching, even-odd oscillations in the photon-number
probability, and negativity of the Wigner
function~\cite{mand2,hill,sch,spe,iss,yue,yam,igo}.  Any of these
features alone, however, does not represent the univocal criterium we are
looking for.  Neither squeezing nor antibunching provides a necessary
condition for nonclassicality~\cite{agar}. The negativity of the
Wigner function, which is well exhibited by the Fock states and the
Schr\"odinger-cat-like states, is absent for the squeezed states.
As for the oscillations in the photon-number probability, some
even-odd oscillations can
be simply obtained by using a statistical mixture of coherent
states~\cite{bond}.  \par Many authors~\cite{agar,kly1,jpa} have
adopted the nonpositivity of the phase-averaged $P$-function
$F(I)={1\over 2\pi}\int _0^{2\pi} d\phi\,P(I^{1/2}e^{i\phi})$ as the
definition for a nonclassical state, since $F(I)< 0$ invalidates
Mandel's semiclassical formula~\cite{mand} of photon counting, i.e., it
does not allow a classical description in terms of a stochastic
intensity. Of course, some states can exhibit a ``weak''
nonclassicality~\cite{jpa}, namely, a positive $F(I)$, but with a
non-positive $P$-function (a relevant example being a coherent state
undergoing Kerr-type self-phase modulation). However, from the point
of view of the detection theory, such ``weak'' nonclassical states
still admit a classical description in terms of having the intensity
probability $F(I)>0$. For this reason, we adopt nonpositivity of
$F(I)$ as the definition of nonclassicality.  
\section{SINGLE-MODE NONCLASSICALITY}
\par The authors of
Refs.~\cite{agar,kly1,jpa} have recognized relations between $F(I)$
and generalized moments of the photon distribution, which, in turn,
can be used to test the nonclassicality. The problem is reduced to an
infinite set of inequalities that provide both necessary and
sufficient conditions for nonclassicality~\cite{kly1}.  In terms of
the photon-number probability $p(n)=\langle n|\hat\rho|n\rangle$ of
the state with density matrix $\hat\rho$, the simplest sufficient
condition involves the following three-point relation for
$p(n)$~\cite{kly1,jpa}
\begin{eqnarray}
B(n)&\equiv&(n+2)p(n)p(n+2) \nonumber
\\&&-(n+1)[p(n+1)]^2<0 \;.\label{p2p}
\end{eqnarray}
Higher-order sufficient conditions involve five-, seven-, \dots, 
$(2k+1)$-point relations, always for adjacent values of $n$.  It is
sufficient that just one of 
these inequalities be satisfied in order to assure the negativity of
$F(I)$. Notice that for a coherent state $B(n)=0$ identically for all
$n$.  \par In the following we show that quantum tomography can be
used as a powerful tool for performing the nonclassicality test in
Eq.~(\ref{p2p}).  For less-than-unity quantum efficiency ($\eta <1$),
we rely on the concept of a ``noisy state'' $\hat\varrho_{\eta}$,
wherein the effect of quantum efficiency is ascribed to the
quantum state itself rather than to the detector. In this model, the
effect of quantum efficiency is treated in a Schr\"odinger-like
picture, with the state evolving from $\hat\varrho$ to
$\hat\varrho_{\eta}$, and with $\eta$ playing the
role of a time parameter.  Such lossy evolution is described by the
master equation
\begin{eqnarray}
\partial_t\hat\varrho (t)=
{\Gamma \over 2}\left\{2
\hat a\hat\varrho (t)\hat a^{\dag}-
\hat a^{\dag}\hat a\hat \varrho(t)- \hat \varrho (t)\hat a^{\dag}\hat a
\right\}\;,\label{mm1}
\end{eqnarray}
wherein $\hat\varrho (t)\equiv\hat\varrho _{\eta }$ 
with $t=-\ln\eta/\Gamma $. 
\par For the nonclassicality test, reconstruction in terms of the
noisy state has many advantages over the true-state reconstruction. In
fact, for nonunit quantum efficiency $\eta <1$ the tomographic method
introduces errors for $p(n)$ which are increasingly large versus $n$,
with the additional limitation that quantum efficiency must be greater
than the minimum value $\eta =0.5$ \cite{rapid,comm}. On the other
hand, the reconstruction of the noisy-state probabilities
$p_{\eta}(n)=\langle n|\hat\rho_{\eta}|n\rangle$ does not suffer such
limitations, and even though all quantum features are certainly
diminished in the noisy-state description, nevertheless the effect of
nonunity quantum efficiency does not change the sign of the
$P$-function, but only rescales it as follows: 
\begin{eqnarray}
P(z)\rightarrow P_{\eta }(z)={1\over\eta}P(z/\eta ^{1/2})
\;.\label{peta}
\end{eqnarray} 
Hence, the inequality (\ref{p2p}) still represents a sufficient 
condition for nonclassicality when the original probabilities 
$p(n)=\langle n|\hat\rho|n\rangle$ are replaced with the noisy-state
probabilities $p_{\eta }(n)=\langle n|\hat\rho_{\eta}|n\rangle$, the
latter being given by the Bernoulli convolution
\begin{eqnarray}
p_{\eta }(n)=\sum _{k=n} ^{\infty} {k\choose n}
\eta ^n (1-\eta)^{k-n} p(k)
\;.\label{bern}
\end{eqnarray}
Hence, when referred to the noisy-state probabilities $p_{\eta }(n)$,
the inequality in Eq.~(\ref{p2p}) keeps its form and simply rewrites
as follows 
\begin{eqnarray}
B_{\eta }(n)&\equiv&(n+2)p_{\eta }(n)p_{\eta }(n+2) \nonumber \\
&&-(n+1)[p_{\eta }(n+1)]^2 < 0 \;.\label{p2pm}
\end{eqnarray}
\par According to
Eq.~(\ref{p2pm}), the quantity $B_{\eta }(n)$ is nonlinear in the
density matrix. This means that $B_{\eta }(n)$ cannot be measured by
averaging a suitable kernel function over the homodyne data, as for
any other observable~\cite{univ}. Hence, in the evaluation of $B_{\eta
}(n)$ one needs to tomographically reconstruct the photon-number
probabilities, using the kernel functions\cite{rapid}
\begin{eqnarray}
K^{(n)}_\eta (x)&=&2\,\kappa ^2\,e^{-\kappa ^2 x^2}
\sum_{\nu=0}^n\frac {(-)^\nu }{\nu !}{n\choose n-\nu }(2\nu +1)!
\kappa ^{2\nu }\nonumber \\& \times & \hbox{Re }\{D_{-(2\nu
+2)}(-2i\kappa x)\}\;,\label{ker}
\end{eqnarray}
where $D_\sigma (z)$ denotes the parabolic cylinder function and 
$\kappa =\sqrt{\eta /(2\eta -1)}$. The true-state probabilities $p(n)$
are obtained by averaging the kernel function in Eq. (\ref{ker}) over
the homodyne data. On the other hand, the noisy-state probabilities
$p_{\eta}(n)$ are obtained by using the kernel function in
Eq. (\ref{ker}) for $\eta=1$, namely without recovering the
convolution effect of nonunit quantum efficiency. Notice that the expression
(\ref{ker}) does not depend on the phase of 
the quadrature. Hence, the knowledge of the phase of the local 
oscillator in the homodyne detector is not needed for the tomographic 
reconstruction, and it can be left fluctuating in a real experiment.  
\par Regarding the estimation of statistical errors, they are
generally obtained by dividing the set of homodyne data into
blocks. However, in the present case, the nonlinear dependence on the
photon number probability introduces a systematic error that is
vanishingly small for increasingly larger sets of data. Therefore, the
estimated value of $B(n)$ will be obtained from the full set of data,
instead of averaging the mean value of the different statistical blocks.
\begin{figure}[htb]
\begin{center}
\epsfxsize=0.6\hsize\leavevmode\epsffile{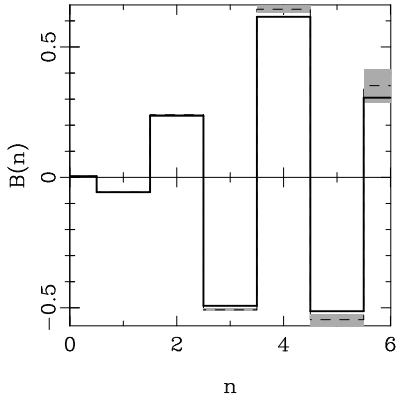}
\end{center}
\caption{Tomographic measurement of $B(n)$ (dashed trace) 
with the respective error bars (superimposed in grey-shade) 
along with the theoretical values (solid trace) 
for a Schr\"odinger-cat state with average photon number 
$\bar n=5$. The quantum efficiency is $\eta =0.8$ and the number of 
simulated experimental data used for the reconstruction is $10^7$.}
\label{f:fig1}
\end{figure}
\begin{figure}[htb]
\begin{center}
\epsfxsize=0.6\hsize\leavevmode\epsffile{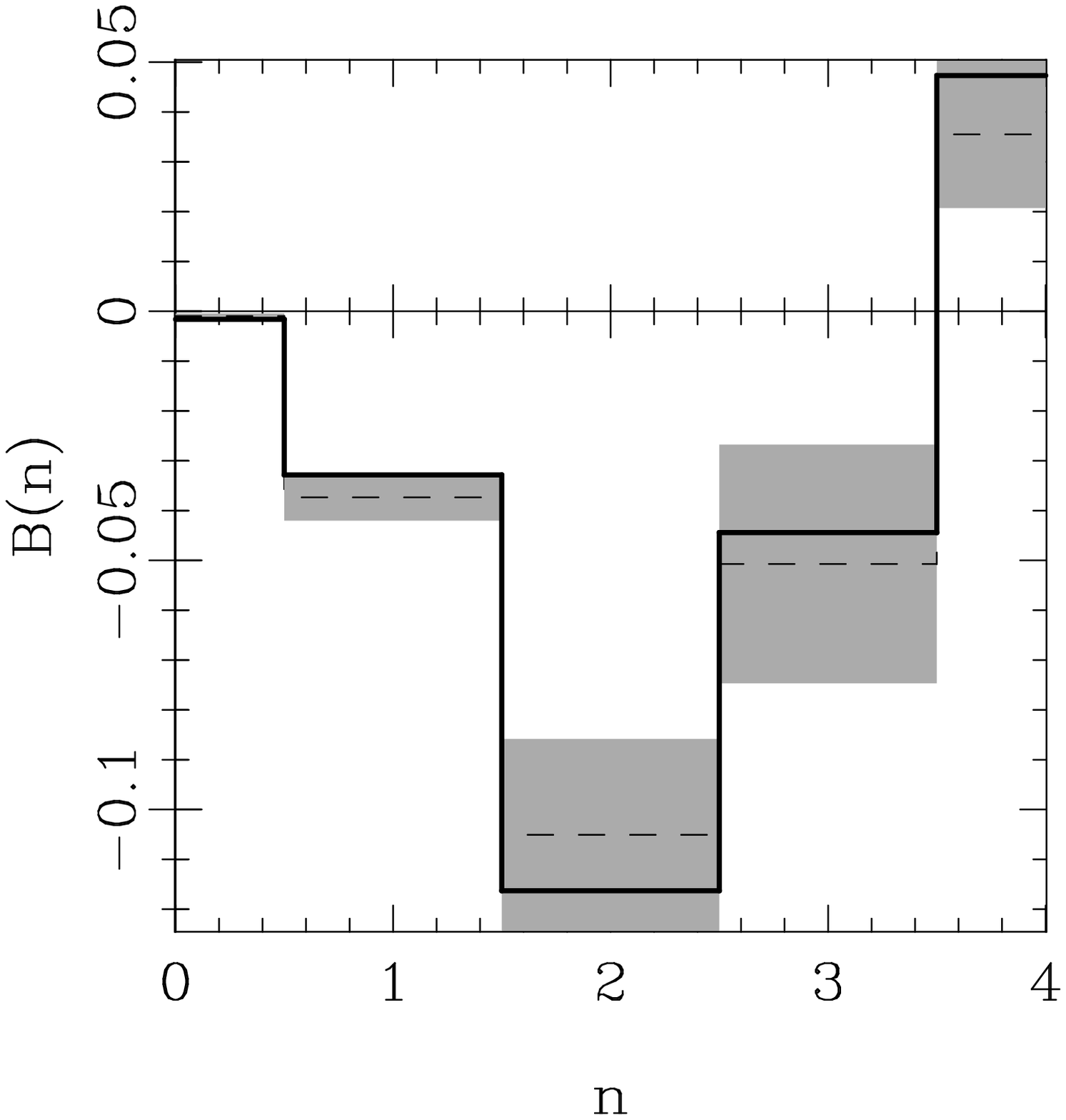}
\end{center}
\caption{Tomographic measurement of $B(n)$ (dashed trace) 
with the respective error bars (superimposed in grey-shade) along with 
the theoretical values (solid trace) 
for a phase-squeezed state with $\bar n=5$ and $\bar n_{\rm sq}=
\sinh^2 r =3$ squeezing photons. The quantum efficiency is $\eta =0.8$ and 
$10^7$ simulated experimental data have been used for the reconstruction.}
\label{f:fig2}
\end{figure}
\par
\begin{figure}[htb]
\begin{center}
\epsfxsize=0.6\hsize\leavevmode\epsffile{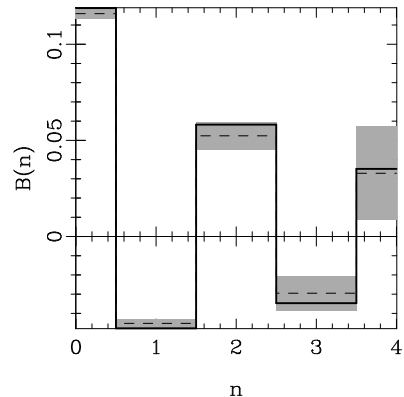}
\end{center}
\caption{Same as in Fig.~\ref{f:fig2}, but for an amplitude-squeezed state.}
\label{f:fig3}
\end{figure}

\begin{figure}[htb]
\begin{center}
\epsfxsize=0.6\hsize\leavevmode\epsffile{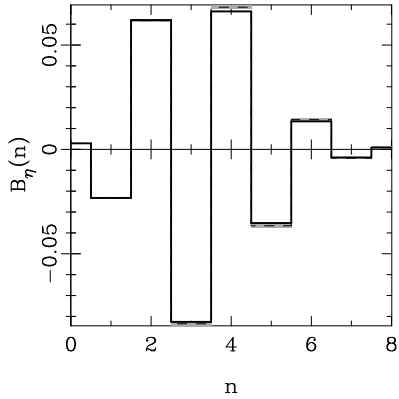}
\end{center}
\caption{Tomographic measurement of $B_{\eta}(n)$ for a 
Schr\"odinger-cat state with $\bar n=5$, degraded by a quantum 
efficiency $\eta =0.8$. The number of simulated experimental data 
is $10^7$.}
\label{f:fig4}
\end{figure}

\begin{figure}[htb]
\begin{center}
\epsfxsize=0.6\hsize\leavevmode\epsffile{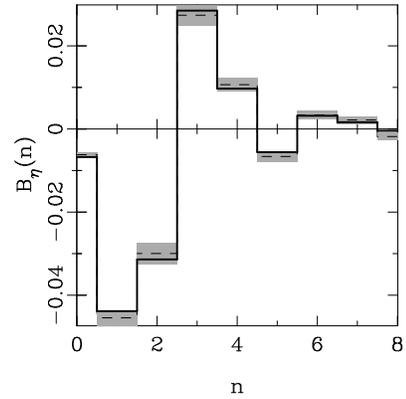}
\end{center}
\caption{Tomographic measurement of $B_{\eta}(n)$ (dashed trace) with the
respective error bars (superimposed in grey-shade) along with the 
theoretical values (solid trace) for a phase-squeezed state, which has
$\bar n=5$ and $\bar n_{\rm sq}=\sinh^2 r =3$ squeezing photons, and which
has been degraded by a quantum efficiency $\eta =0.8$. For the
reconstruction a sample of $10^7$ simulated experimental data have
been used.} 
\label{f:fig5}
\end{figure}

\begin{figure}[htb]
\begin{center}
\epsfxsize=0.6\hsize\leavevmode\epsffile{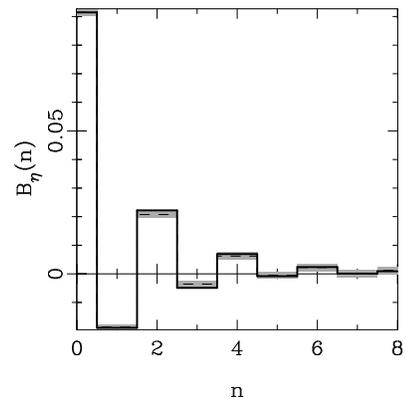}
\end{center}
\caption{Same as in Fig.~\ref{f:fig5}, but for an amplitude-squeezed state.}
\label{f:fig6}
\end{figure}

\begin{figure}[htb]
\begin{center}
\epsfxsize=0.6\hsize\leavevmode\epsffile{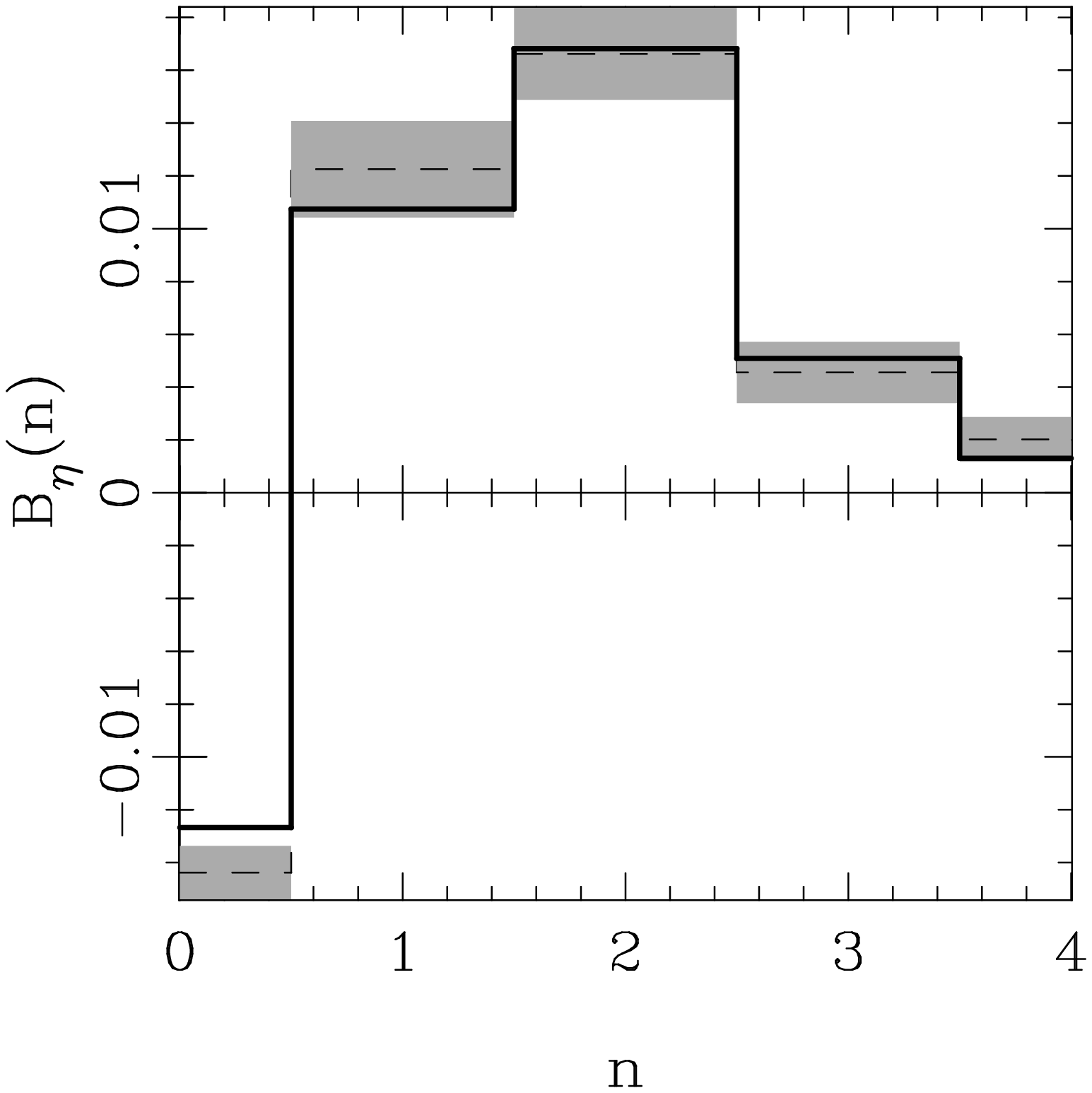}
\end{center}
\caption{Same as in Fig.~\ref{f:fig5}, but here for a quantum efficiency 
of $\eta =0.4$, and a sample of $5\times 10^7$ simulated experimental
data.} 
\label{f:fig7}
\end{figure}

 In Figs.~\ref{f:fig1}--\ref{f:fig7} we present some numerical
results that are obtained by a Monte-Carlo simulation of a quantum
tomography experiment. The nonclassicality criterium is tested either
on a Schr\"odinger-cat state $|\psi (\alpha )\rangle \propto
(|\alpha\rangle +|-\alpha\rangle )$ or on a squeezed state
$|\alpha,r\rangle\equiv D(\alpha )S(r)|0\rangle $, wherein $|\alpha
\rangle$, $D(\alpha )$, and $S(r)$ denote a coherent state with
amplitude $\alpha $, the displacement operator $D(\alpha )=e^{\alpha
\hat a^{\dag }-\bar\alpha \hat a}$, and the squeezing operator
$S(r)=e^{r(\hat a^{\dag 2}-\hat a^2)/2}$,
respectively. Figs.~\ref{f:fig1}--\ref{f:fig3} show
tomographically-obtained values of $B(n)$, with the respective error
bars superimposed, along with the theoretical values for a
Schr\"odinger-cat state, for a phase-squeezed state ($r>0$), and for
an amplitude-squeezed state ($r<0$), respectively. 
For the same set of states the results for $B_{\eta}(n)$
[cf.\ Eq.~(\ref{p2pm})] obtained by tomographic reconstruction of the
noisy state are reported in Figs.~\ref{f:fig4}--\ref{f:fig6}. Let us
compare the statistical errors that affect the two measurements,
namely, those of $B(n)$ and $B_{\eta }(n)$ on the original and the
noisy states, respectively. In the first case
(Figs.~\ref{f:fig1}--\ref{f:fig3}) the error increases with $n$,
whereas in the second (Figs.~\ref{f:fig4}--\ref{f:fig6}) it remains
nearly constant, albeit with less marked oscillations in $B_{\eta
}(n)$ than those in $B(n)$.  Fig.~\ref{f:fig7} shows
tomographically-obtained values of $B_{\eta}(n)$ for the
phase-squeezed state (cf.\ Fig.~\ref{f:fig5}), but for a lower quantum
efficiency $\eta =0.4$. Notice that, in spite of the low quantum
efficiency, the nonclassicality of such a state is still
experimentally verifiable, as $B_{\eta }(0)< 0$ by more than five
standard deviations.  In contrast, for coherent states one obtains
small statistical fluctuations around zero for all $n$.
We remark that the simpler test of checking for
antibunching or oscillations in the photon-number probability in the
case of the phase-squeezed state considered here (Figs.~\ref{f:fig2},
\ref{f:fig5}, and \ref{f:fig7}) would not reveal the nonclassical
features of such a state.
\section{TWO-MODE NONCLASSICALITY}
\par Quantum homodyne tomography can also
be employed to test the nonclassicality of two-mode states. For a
two-mode state nonclassicality is defined in terms of nonpositivity of
the following phase-averaged two-mode $P$-function~\cite{jpa}:
\begin{eqnarray}
F(I_1,I_2,\phi )={1\over 2\pi}\int_0^{2\pi}d\phi_1\,
P(I_1^{1/2}e^{i\phi _1},I_2^{1/2}e^{i(\phi _1+\phi )})
\;.\label{non2} 
\end{eqnarray}
In Ref.~\cite{jpa} it is also shown that a sufficient condition for 
nonclassicality is
\begin{eqnarray}
C=\langle (\hat n_1 -\hat n_2)^2\rangle-
(\langle \hat n_1 -\hat n_2\rangle )^2-
\langle \hat n_1 +\hat n_2\rangle <0
\;,\label{n2} 
\end{eqnarray}
where $\hat n_1$ and $\hat n_2$ are the photon-number operators of the 
two modes. 
\par
A tomographic test of the inequality in Eq.~(\ref{n2}) can be performed by 
averaging the kernel functions for the operators in the ensemble averages in 
Eq.~(\ref{n2}) over the two-mode homodyne data. For the normal-ordered 
field operators one can use the Richter formula in
Ref.~\cite{richter}, namely 
\begin{eqnarray}
{\cal R}[a^{\dag n}a^m](x,\phi)=e^{i(m-n)\phi}\,\frac{H_{n+m}(\sqrt{2\eta
}x)}{\sqrt{(2\eta)^{n+m}}{n+m\choose n}}\;,\label{ric}
\end{eqnarray}
$H_n(x)$ denoting the Hermite polynomial and $\phi$ being the phase
of the fiels with respect to the local oscillator of the homodyne
detector. Again, as for the kernel function in Eq. (\ref{ker}), the
value $\eta=1$ is used to reconstruct the ensemble averages of the
noisy state $\hat\rho_{\eta}$. Notice that for $n=m$ Eq. (\ref{ric})
is independent on the phase $\phi$, and hence no phase knowledge is
needed to reconstruct the ensemble averages in Eq. (\ref{n2}).  
As an example, we consider the twin-beam state at the output of a
nondegenerate parametric amplifier 
\begin{eqnarray}
|\chi\rangle \equiv(1-|\lambda |^2)\sum _{n=0}^{\infty }\lambda ^n 
|n\rangle\otimes|n\rangle\;,
\label{twb} 
\end{eqnarray}
where $|n\rangle\otimes|n\rangle$ denotes the joint eigenvector of the
number operators of the two modes with equal eigenvalue $n$, and the
parameter $\lambda $ is related to the gain $G$ of the amplifier by
the relation $|\lambda |^2=1-G^{-1}$.  The theoretical value of $C$
for the state in Eq.~(\ref{twb}) is
$C=-2|\lambda|^2/(1-|\lambda|^2)<0$.  A tomographic reconstruction of
the twin-beam state in Eq.~(\ref{twb}) is particularly facilitated by
the self-homodyning scheme, as shown in Ref.~\cite{self}.  With regard
to the effect of quantum efficiency $\eta <1$, the same argument still
holds as for the single-mode case: one can evaluate $C_{\eta}$ for the
twin-beam state that has been degraded by the effect of loss.  In this
case, the theoretical value of $C_\eta$ is simply rescaled to $C_\eta
=-2\eta ^2 |\lambda|^2/(1-|\lambda|^2)$.
\begin{figure}[htb]
\begin{center}
\epsfxsize=0.6\hsize\leavevmode\epsffile{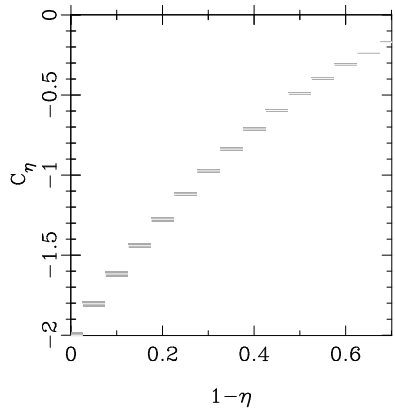}
\end{center}
\caption{Tomographic measurement of $C_{\eta}$ as defined in
Eq.~(\ref{n2}) and modified by the quantum efficiency for the
twin-beam state in Eq.~(\ref{twb}).  The respective error bars are
shown in the grey shade and $|\lambda|^2=0.5$ corresponding to an
average of 2 total photons.
The results are shown for various values of the quantum efficiency
$\eta $ (in steps of 0.05) and for each value of $\eta $ the number of
simulated 
data is $4\times 10^5$.}
\label{f:fig9}
\end{figure}
\par In Fig.~\ref{f:fig9} we report $C_{\eta }$ {\em vs.} 
$1-\eta $, $\eta $ ranging from 1 to 0.3 in steps of 0.05, 
for the twin-beam state in Eq.~(\ref{twb}) with $|\lambda |^2=0.5$, 
corresponding to the total average photon number equal to 2. 
The values of $C_{\eta }$ result from a Monte-Carlo simulation of a
homodyne tomography experiment with a sample of $4\times 10^5$
data, using the theoretical joint homodyne probability of the state
$|\chi\rangle$ 
\begin{eqnarray}
p_{\eta }(x_1,x_2,\phi _1,\phi _2)= {2\exp\left[
-{{(x_1+x_2)^2}\over{d^2_{z}+4\Delta^2_{\eta}}}-
{{(x_1-x_2)^2}\over{d^2_{-z }+4\Delta^2_{\eta}}}\right]
\over{\pi\sqrt{(d^2_{z }+4\Delta^2_{\eta})(d^2_{-z }+4
\Delta^2_{\eta})}}}\;,\label{pxy}
\end{eqnarray} 
with
\begin{eqnarray}
&&z =e^{-i(\phi _1+\phi _2)}\Lambda\;,\nonumber\\
&&d^2_{\pm z }={{|1\pm z |^2}\over{1-|z |^2}}\;,\nonumber\\
&&\Delta ^2_{\eta }=\frac{1-\eta}{4\eta}\;,
\end{eqnarray}
$\phi _1$ and $\phi _2$ denoting the phases of the two modes relative
to the respective local oscillator.
Notice that the nonclassicality test in terms of the noisy state gives 
values of $C_{\eta }$ that are increasingly near the classically
positive region for decreasing quantum efficiency $\eta $. However,
the statistical error remains constant and is sufficiently small to 
allow recognition of the nonclassicality of the twin-beam state in 
Eq.~(\ref{twb}) up to $\eta =0.3$.
\section{CONCLUSIONS}
We have shown that quantum homodyne tomography allows one to 
perform nonclassicality tests for various single- and two-mode radiation 
states, even when the quantum efficiency of homodyne detection is rather low. 
The method involves reconstruction of the photon-number probability or of 
some suitable function of the number operators pertaining to the 
noisy state, namely, the state degraded by the less-than-unity
quantum efficiency. The noisy-state reconstruction is 
affected by the statistical errors; however, they are sufficiently small
that the nonclassicality of the state can be tested even for low values of 
$\eta $. For the cases considered in this paper, we have shown that the 
nonclassicality of the states can be proven (deviation from classicality 
by many error bars) with $10^5$--$10^7$ 
homodyne data. Moreover, since the knowledge of the phase of the local 
oscillator in the homodyne detector is not needed for the tomographic 
reconstruction, it can be left fluctuating in a real experiment. 
Hence, we conclude that the proposed nonclassicality test should be easy 
to perform experimentally.
\section*{Acknowledgments}
This work is supported by the Italian Ministero dell'Universit\'a e
della Ricerca Scientifica e Tecnologica under the program {\em 
Amplificazione e rivelazione di radiazione quantistica} 

\end{document}